\documentclass[graybox]{svmult}

\pdfoutput=1
\usepackage{amsmath,amssymb}
\usepackage{setspace}
\usepackage{mathptmx}       
\usepackage{helvet}         
\usepackage{courier}        
\usepackage{type1cm}        
\usepackage{makeidx}         
\usepackage{graphicx}        
\usepackage{multicol}        
\usepackage[bottom]{footmisc}

\usepackage{sidecap}
\usepackage{listings}
\makeindex             

\begin{document}

\title*{Triple cascade behaviour in QG and drift turbulence and generation of zonal jets}
\author{Sergey Nazarenko, Brenda Quinn}
\institute{Sergey Nazarenko \at Mathematics Institute, University of Warwick, Gibbet Hill
Road, Coventry CV4 7AL, UK, \email{S.V.Nazarenko@.warwick.ac.uk}
\and Brenda Quinn \at Mathematics Institute, University of Warwick, Gibbet Hill
Road, Coventry CV4 7AL, UK \email{B.E.Quinn@warwick.ac.uk}}
%
%
\maketitle

\abstract{
We study quasi-geostrophic turbulence and plasma drift turbulence within the Charney-Hasegawa-Mima (CHM) model.
We focus, theoretically and using numerical simulations, on  conservation of {\em zonostrophy} and on its role in the formation of the zonal jets. The zonostrophy invariant
was first predicted in \cite{perm,BNZ_invariant}   in two special cases -- large-scale turbulence
and anisotropic turbulence.
Papers \cite{perm,BNZ_invariant}   also predicted that the three invariants,
energy, enstrophy and zonostrophy, will cascade anisotropically into non-intersecting sectors
in the $k$-space, so that the energy cascade is "pushed"
into the large-scale  zonal scales.
In the present paper, we  consider the scales much
 less than the Rossby deformation radius
  and generalise the
 Fj{\o}rtoft argument of  \cite{perm,BNZ_invariant}  to find the directions of the
 three cascades in this case.
For the first time, we demonstrate numerically  that zonostrophy is well conserved by the CHM model,
and that the energy, enstrophy and zonostrophy cascade as prescribed by the
Fj{\o}rtoft argument
if the nonlinearity is sufficiently weak. Moreover, numerically we observe that zonostrophy is conserved
surprisingly well at late  times and the triple-cascade picture is rather
accurate even if the initial nonlinearity is strong.}


\section{Introduction and the model}
\label{sec:intro}

Zonal jets are prominent features in  geophysical fluids, e.g. atmospheres of  Jupiter and Saturn \cite{simon,saturn,Galperin1} and the earths' atmosphere \cite{atmospheric_jets,gill_82} and oceans \cite{gill_82,Galperin1,maximenko}.
Zonal jets have also been observed in fusion plasmas  \cite{diamond}.
Zonal jets are important because they can suppress the small-scale turbulence and block the transport
 in both geophysical settings \cite{james}
and in plasmas \cite{BNZ90,wagner,diamond}.

Two main zonal jet generation mechanisms considered in the literature are the modulational
instability \cite{lorentz,gill,mima_lee,MN94,smolyakov,onishchenko,smolyakov2,MI-ours} and
the anisotropic inverse cascade
\cite{Williams78,Rhines,Rhines79,BNZ90}.
The inverse cascade mechanism brings the energy from initial small-scale turbulence to the large-scale
 zonal flows in a step-by-step (local in the scale space) transfer mechanism
 similar to the inverse cascade  in 2D Navier-Stokes turbulence \cite{fjortoft,kraichnan67}.
 On the other hand, the modulational instability
 brings energy to the large zonal scales directly, skipping the intermediate scales.
 It appears to be more relevant if the  small-scale turbulence has a narrow
initial spectrum, whereas the cascade picture is more accurate when the initial
spectrum is broad. We will report on our recent analytical and numerical studies of the modulational instability elsewhere \cite{MI-ours}, whereas the present paper will focus on the turbulent cascades.

The mechanism for  an inverse cascade in geophysical quasi-geostrophic (QG) turbulence
and in plasma drift turbulence  is quite
similar to the one of the 2D Navier-Stokes turbulence \cite{fjortoft,kraichnan67}, but
the presence of the beta-effect (non-uniform rotation of geophysical fluids or plasma inhomogeneity)
makes this cascade anisotropic leading to condensation into
large-scale zonal flows rather than round vortices  \cite{Williams78,Rhines,Rhines79}.

In this paper we will follow the approach put forward in papers \cite{perm,BNZ_invariant} which
is most relevant (and asymptotically rigorous) when the QG/drift turbulence is weak.
In this case the turbulence is dominated by waves which are involved in triad interactions.
These three-wave interactions are shown to conserve an additional positive
quadratic invariant. This invariant and  the other two quadratic invariants, the energy and the potential
enstrophy, are involved in a triple cascade process, which can be described via an
argument similar to the standard Fj{\o}rtoft argument originally developed
for the 2D Navier-Stokes turbulence \cite{fjortoft}. It was found that each of the invariants is
forced by the other two to cascade into its own anisotropic sector of scales and, in particular,
the energy is forced to cascade to long zonal scales.
Considering its important role in the zonation process, hereafter we will label the extra invariant
as {\em zonostrophy}.

On the other hand, work
\cite{perm,BNZ_invariant} was limited to considering either very large scales (longer than the
Rossby deformation radius or the Larmor radius) or to the scales which are already
anisotropic and are close to zonal. Besides, the conservation of the extra invariant is based
on the weakness of nonlinearity and on the randomness of phases (conditions of validity of the wave
kinetic equation), which, even when present initially, can break down later on during the zonation process.
Thus, numerical checks of robustness of the zonostrophy conservation  were needed, and they
have not been done until the work reported in the present paper.

Soon after papers \cite{perm,BNZ_invariant}, the zonostrophy invariant was generalized to
the whole of the $k$-space in \cite{balk-gen}.
 This was a significant achievement because the extra invariant of such a kind appeared to be
 unique for Rossby/drift systems and is not observed in any other known nonlinear wave model
 \cite{balk-siam}. Besides, its conservation has revealed interesting geometrical properties
 of the wave dispersion   relation \cite{balk-siam}.
 Unfortunately, the general expression for zonostrophy appeared to have
 a form for which the Fj{\o}rtoft argument cannot be used (not scale invariant, not sign-definite).
 However, an alternative zonation argument was put forward in \cite{balk-zonation}.

 In the present paper we will focus on the special case when the scales are much smaller than
 the deformation (Larmor) radius, which is the most important and frequently considered limit
 (at least in the GFD context). Taking the respective limit in the general zonostrophy
 expression obtained in \cite{balk-gen},  we obtain the zonostrophy expression
 for such small-scale turbulence and show that it is positive and scale invariant.
 The latter observation is extremely important because it means that we can once again
 apply the generalized  Fj{\o}rtoft argument developed in \cite{perm,BNZ_invariant}, which is done in the present
 paper. Note that the Fj{\o}rtoft argument of \cite{perm,BNZ_invariant} is somewhat preferential
 over the argument presented in \cite{balk-zonation} because it predicts not only zonation but
 also the anisotropic $k$-space flow paths of the three invariants during the zonation process.

Having obtained these analytical predictions, we then proceed to direct numerical simulations (DNS)
of the QG/drift turbulence to (i) test the conservation of the zonostrophy for different levels
of initial nonlinearity, and (ii) test predictions of the generalized  Fj{\o}rtoft argument by tracking in time the transfer path of the three invariants in the $k$-space. As a result, we confirm conservation of zonostrophy
and its important role in directing the energy to the zonal jet scales.

\section{Charney-Hasegawa-Mima model}
\label{sec:CHM}

The reason why geophysical quasi-geostrophic (QG) flows
and plasma drift turbulence are often mentioned together (in particular when discussing the zonal jet formation)
is that some basic linear and nonlinear properties in these two systems can be described by the
same PDE, the Charney-Hasegawa-Mima
 (CHM) equation \cite{charney,hasegawa_mima}:
\begin{equation}
\label{eq-CHM}
\partial_t \Delta \psi  + \beta \partial_x \psi +
(\partial _x \psi) \partial_y \Delta \psi - (\partial _y \psi) \partial_x \Delta \psi = 0,
\end{equation}
where
$\psi$ is the streamfunction, $\beta$ is a constant proportional to the gradient of the horizontal rotation frequency or of the plasma
density in the GFD and plasma contexts respectively.
In the GFD context, the $x$-axis is in the west-east  and the $y$-axis
is along the south-north directions respectively.
In plasmas, the $y$-axis is along the plasma density gradient and the $x$-axis, of course, is transverse to this direction.
Here, we consider only small-scale turbulence with scales much
smaller than the Rossby deformation radius in GFD and the ion Larmor radius in  plasma contexts (this has already been
taken into account in the CHM model,~(\ref{eq-CHM})).

Let us put our system in a periodic square box with side length $L$ and
introduce the Fourier transform of the streamfunction,
$$\psi_\textbf{k} = \frac{1}{L^2} \int \psi(\textbf{x}) e^{-i \textbf{k} \cdot \textbf{x}} \, d \textbf{x},$$
where  $\textbf{k}=(k_x,k_y)$ is a wavevector in the 2D plane.
In $k$-space the  CHM equation (\ref{eq-CHM}) becomes:
\begin{eqnarray}
\partial_t \hat \psi_\textbf{k} = i\, \omega_\textbf{k}\, \hat \psi_\textbf{k}
+ \frac{1}{2} \sum_{
\textbf{k}_1 + \textbf{k}_2=\textbf{k}} T(\textbf{k},\textbf{k}_1,\textbf{k}_2)\,
\hat  \psi_{\textbf{k}_1}\, \hat  \psi_{\textbf{k}_2}\, ,
  \label{eq-CHMk}
\end{eqnarray}
where
\begin{equation}
\label{eq-RossbyDispersion}
\omega_\textbf{k} = -\frac{\beta k_x}{k^2 }
\end{equation}
is the dispersion relation for the frequency of linear waves (Rossby or drift waves
in the GFD and plasma contexts respectively),  $k = \left| \textbf{k} \right|$, and
\begin{equation}
\label{eq-CHMInteractionCoefficient}
T(\textbf{k},\textbf{k}_1,\textbf{k}_2) = -\frac{\left(\textbf{k}_1 \times \textbf{k}_2\right)_z (k_1^2-k_2^2)}{k^2 }
\end{equation}
is a nonlinear interaction coefficient.

\section{Conservation of energy and enstrophy} \label{}

It is well known that the CHM equation,~(\ref{eq-CHM})
conserves the energy and the enstrophy which in physical space are defined respectively
as
\begin{equation}
\label{eq:energy_x}
E = \frac{1}{2} \int (\nabla\psi)^2 \hspace{1mm} d \textbf{x}
\end{equation}
and
\begin{equation}
\label{eq:enstrophy_x}
\Omega = \frac{1}{2} \int (\triangle \psi)^2  \hspace{1mm} d \textbf{x}.
\end{equation}
In Fourier space these conserved quantities can be expressed in terms of the wave action
\begin{equation}
\nonumber
n(\textbf{k}) = \frac{k^4 |\hat \psi_k|^2}{2 \beta k_x}.
\end{equation}
We have
\begin{equation}
\label{eq:energy_k}
E =  \int |\omega_\textbf{k}| n_\textbf{k} \hspace{1mm} d \textbf{k}
\end{equation}
and
\begin{equation}
\label{eq:enstrophy_k}
\Omega =  \int k_x n_\textbf{k} \hspace{1mm} d \textbf{k}.
\end{equation}

\section{Conservation of zonostrophy.}

The energy and the enstrophy are {\em exact} invariants of the
small-scale CHM model (\ref{eq-CHM}).  {\em Zonostrophy} is an exact invariant of the {\em kinetic equation}:
\begin{eqnarray}
\label{kineq}
	\dot n_\textbf{k}
=
	\int\hspace{-0.5mm} &  \left| V_{12k} \right|^2 \delta(\textbf{k}_1 + \textbf{k}_2 -\textbf{k}) \delta(\omega(\textbf{k}_1) + \omega(\textbf{k}_2) -\omega(\textbf{k})) \times
\nonumber \\
\nonumber
	& \left[ n(\textbf{k}_1) n(\textbf{k}_2) - 2 n(\textbf{k}) n(\textbf{k}_1) \hspace{.5mm} \mathrm{sign} 
-  n(\textbf{k}) n(\textbf{k}_2) \hspace{.5mm} \mathrm{sign} \hspace{.5mm} (k_x k_{2x})
 \right]
\ d \textbf{k}_1 d \textbf{k}_2,
\end{eqnarray}
where
\begin{equation}
\nonumber
V_{12k} =  |k_x k_{1x} k_{2x}|^{1/2} \left(\frac{k_{1y}}{k_{1}^2} + \frac{k_{2y}}{k_{2}^2} - \frac{k_{y}}{k^2}\right).
\end{equation}
Thus, it should be clear that the zonostrophy is only proven to be a conserved quantity under the same conditions
for which the kinetic equation (\ref{kineq}) is valid, namely {\em weak nonlinearity} and {\em random phases}.
It is presently unknown if the zonostrophy conservation extends to a broader range of situations or not.
Thus, the numerical tests of zonostrophy conservation are crucial, and this is one of the aims of the
present work.

In terms of the wave action spectrum, zonostrophy
 $Z$ can be written as
\begin{eqnarray}
\label{eq:zonostrophy_k}
Z =  \int \zeta_\textbf{k} n_\textbf{k} \hspace{1mm} d \textbf{k},
\end{eqnarray}
where function $\zeta_\textbf{k}$ is the density of zonostrophy in the $k$-space which
satisfies the triad resonance condition
\begin{equation}
\label{eq:eta-res}
\zeta(\textbf{k}) = \zeta(\textbf{k}_1) +\zeta(\textbf{k}_2)
\end{equation}
for all wavevectors $\textbf{k}, \textbf{k}_1$ and  $\textbf{k}_2$ which lie
on the resonant surface given by the solutions of the wavevector
and frequency resonance conditions,
\begin{equation}
\label{eq:eta1}
\textbf{k}= \textbf{k}_1 +\textbf{k}_2
\end{equation}
and
\begin{equation}
\label{eq:eta2}
\omega(\textbf{k}) = \omega(\textbf{k}_1) +\omega(\textbf{k}_2).
\end{equation}
Expressions for $\zeta_\textbf{k}$ were first found in  \cite{perm,BNZ_invariant}  in the special cases of large-scale
turbulence ($\rho k \ll 1$  where $\rho$ is the Rossby deformation radius in GFD or ion Larmor radius in plasmas)
and for  anisotropic  turbulence ( $k_x \ll k_y$), after which a general expression was found for all
$\textbf{k}$'s in \cite{balk-gen}. This expression is
\begin{equation}
\label{eq:eta-gen}
\zeta(\textbf{k}) =\arctan \frac{k_y + k_x \sqrt{3}}{\rho k^2} - \arctan \frac{k_y - k_x \sqrt{3}}{\rho k^2}.
\end{equation}
This expression can change sign and it is not a scale-invariant function of the wavevector
components, which means that one cannot use the generalized  Fj\o rtoft argument of
\cite{perm,BNZ_invariant}  to find the cascade directions in such a general case.

To find zonostrophy in the special case of small-scale turbulence (which is considered in the present
paper) we have to take the limit $\rho k \to \infty$ in the general expression (\ref{eq:eta-gen}).
It turns out that naively taking limit leads to an already known integral, the
energy and one has to go to further orders in the Taylor expansion of
(\ref{eq:eta-gen}) until we reach an expression which is independent of $E$ and $\Omega$.
Interestingly, such an independent invariant appears only in the fifth order, and the derivation
details are given in Appendix A. The result is
\begin{equation}
\label{eq:eta-ss}
\tilde \zeta = - \lim_{\rho \to \infty}\frac{5 \rho^5}{8 \sqrt{3}} (\zeta - 2 \sqrt{3}  \omega/\beta \rho) =
\frac{{k_x}^3}{{k}^{10}}({k_x}^2 + 5{k_y}^2).
\end{equation}
The integral (\ref{eq:zonostrophy_k}) with
the density (\ref{eq:eta-ss}) is an exact invariant of the kinetic equation  (\ref{kineq}) and thus
it is an {\em approximate} invariant of the small-scale CHM equation (\ref{eq-CHM}).
Later on we will examine numerically the precision with which this invariant is conserved.

Expression (\ref{eq:eta-ss}) for the new invariant in the small-scale limit, $\rho k \gg 1$,
 allows us to explicitly see that the invariant's
density is strictly positive and scale-invariant,
which is good news because once again one can use the generalized
Fj\o rtoft argument of
\cite{perm,BNZ_invariant}  to find the cascade directions of the three invariants:
the energy, the enstrophy and the zonostrophy.

\section{Triple cascade behaviour.}

\subsection{Dual cascades in 2D Navier-Stokes turbulence.}

Before we present the generalized
Fj\o rtoft argument, we will remind the reader of the classical Fj\o rtoft argument
leading to the prediction of the dual cascade behaviour in 2D Navier-Stokes turbulence
\cite{fjortoft}. This will be instructive because by now there exists
several versions of such an argument and the one we use here
is not necessarily the most familiar.

\begin{figure}[h!]
\begin{centering}
\includegraphics[width=0.65\textwidth]{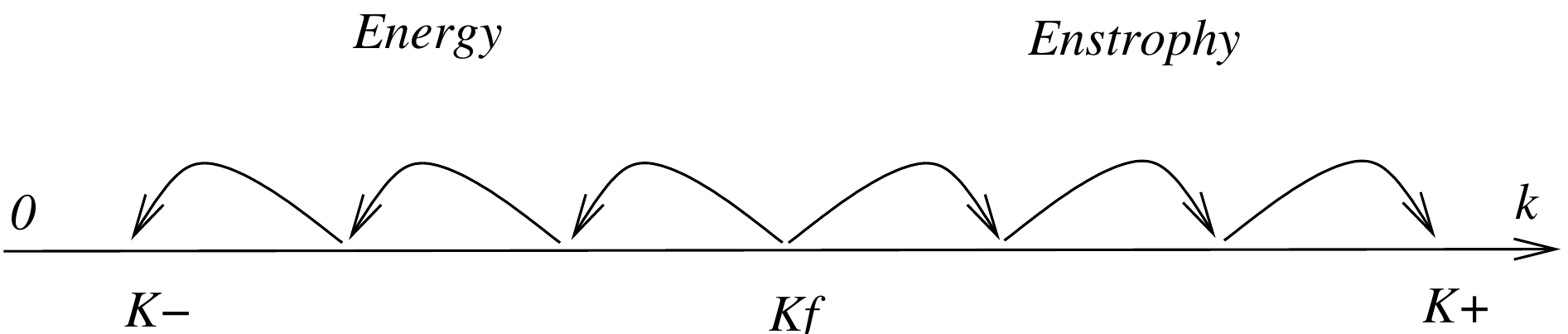}
\caption{\label{fig:dual-casc} Dual cascade behaviour in 2D Navier-Stokes turbulence.}
\end{centering}
\end{figure}

Let us consider 2D turbulence described by the Navier-Stokes equations
excited at some forcing scale $\sim k_0$ and dissipated at very large ($\sim k_- \ll k_0$) and at very small scales ($\sim k_+ \gg k_0$), see
Fig.~\ref{fig:dual-casc}.
The conservative ranges between the forcing scale and the dissipation scales,
 $k_- \ll k \ll k_0 $ and $ k_0 \ll k \ll  k_+ $
are called the inertial ranges.
The two conserved quantities in this case, in the absence of forcing and dissipation,
 are the energy and the enstrophy which
are given by the same expressions as for the small-scale CHM model, namely
by (\ref{eq:energy_x}) and  (\ref{eq:enstrophy_x}) in the $x$-space
and by
(\ref{eq:energy_k}) and  (\ref{eq:enstrophy_k}) in the $k$-space.
In the presence  of forcing and dissipation in steady-state turbulence,
the rate of production of the energy and the enstrophy
 by forcing must be exactly the same as the dissipation rates,
$\varepsilon$ and $\eta$ respectively. Our task now is to determined
where exactly $E$ and $\Omega$ are dissipated, at
$\sim k_- $ or at $\sim k_+ $.

First, we note that the $k$-space densities of $E$ and $\Omega$ differ by a factor of
$k^2$, and therefore the energy dissipation rate
$\varepsilon$ is related to the enstrophy dissipation rate $\eta$ as $\eta \sim {k_0}^2 \varepsilon$.
Now, let us suppose {\em ad absurdum} that at $\sim k_+ $ the energy is dissipated at a
rate comparable to the rate of production at the forcing scales, i.e. $\sim  \varepsilon$.
But this would mean that the enstrophy would be dissipated at the rate $\sim k_+^2 \varepsilon$
which is impossible in steady state since this amount greatly exceeds the
rate  of the  enstrophy production $\eta \sim \sim {k_0}^2 \varepsilon$.
Thus we conclude that most of the energy must be dissipated at
the scales $\sim k_- $, i.e. that the energy cascade is {\em inverse} (with respect to its
direction in 3D turbulence).

Similarly, assuming {\em ad absurdum}  that most of the enstrophy is dissipated
at $\sim k_- $ would also lead to a conclusion that the amount of energy
dissipated is much greater than the energy produced, which is impossible.
Therefore, the enstrophy cascade is direct, i.e. it is dissipated at wavenumbers
$\sim k_+ $ which are greater than the forcing scale $k_0$.

\subsection{Triple cascades in CHM turbulence.}

From the Fj\o rtoft argument presented above, the reader
should notice that the quantities that determine the cascade directions
are the $k$-space densities of the invariants or, more precisely, the scaling of the ratios of these
densities with $k$.
It is also important that these $k$-space densities are positive (as is the case for
$E, \Omega$ and $Z$ in our situation, see (\ref{eq:energy_k}), (\ref{eq:enstrophy_k})
and (\ref{eq:eta-ss})) because otherwise a large positive and a large
negative amount of the same invariant could be produced in different corners of the $k$-space,
with the net result zero and the  Fj\o rtoft argument would not work.
 Further, it is important that all of the invariants have the same scaling
with respect to the turbulence intensity so that their ratios are functions of ${\bf k}$ and not of
the turbulent intensity. This condition is satisfied, for both  Navier-Stokes and in
the CHM model since $E, \Omega$ and $Z$ are linear in $n_k$, see
(\ref{eq:energy_k}), (\ref{eq:enstrophy_k})
and (\ref{eq:eta-ss}).

The ratio of the  $k$-space densities of $\Omega$ and $E$ is $k^2$ which
allows us to conclude that $\Omega$ must go to $k \gg k_0$ and $E$ must cascade
to $k \ll k_0$. In other words, each of the invariants cascades
to the scales where its density is dominant  over the density of the other invariant.
This argument is easily generalizable to the CHM model and the respective
three invariants, $E, \Omega$ and $Z$.
Now we have three invariants, and the cascade picture would necessarily be
anisotropic (it is impossible to divide the 2D ${\bf k}$-space into three non-intersecting
cascade regions in an isotropic way).

\begin{figure}[h!]

\includegraphics[width=0.65\textwidth,angle=-90]{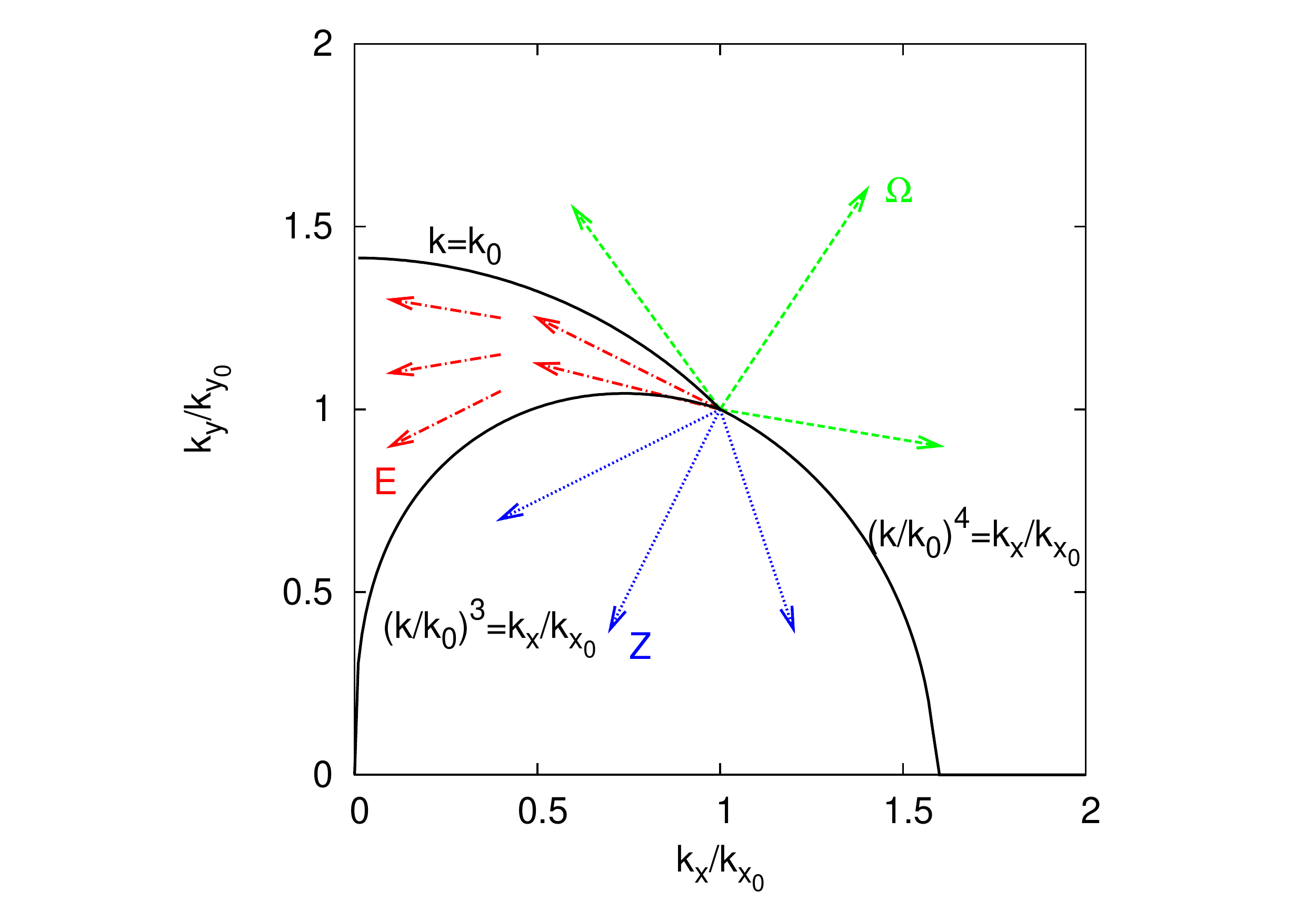}
\begin{centering}
\caption{\label{fig:triple_sketch} Non-intersecting sectors for triple cascade as predicted by the generalized Fj\o rtoft argument.}
\end{centering}
\end{figure}

Let us suppose that turbulence is produced near ${\bf k}_0 = (k_{0x}, k_{0y})$
and it can be dissipated only in regions which are separated in scales from the forcing
scale, i.e. either at $k \gg k_0$ (short scales), or at  $k_{x} \ll k_{0x}$ (nearly zonal scales),
or at  $k_{y} \ll k_{0y}$ (nearly meridional scales), see
Fig.~(\ref{fig:triple_sketch}).
Then each of the invariants (e.g. $E$),
must cascade
to the scales where its density is dominant over the densities of the other two invariants (e.g. $\Omega$ and $Z$).
The boundaries between the  cascading ranges lie on the curves in the
$k$-space where the ratios of the different invariant densities (taken pairwise)  remain constant
(equal to the respective initial values).

We note that because $k^2 \le {k_x}^2 + 5{k_y}^2  \le 5 k^2$,
and because because the Fj\o rtoft argument operates only with the strong inequalities ($\gg$ and $\ll$ rather
than $>$ and $>$), we can replace the zonostrophy density
(\ref{eq:eta-ss}) with a simpler expression,
\begin{equation}
\label{eq:eta-ss-f}
\tilde \zeta_\textbf{k} \sim
\frac{{k_x}^3}{{k}^{8}}.
\end{equation}

Thus we have for the boundaries between the different cascade sectors, see
Fig.~(\ref{fig:triple_sketch}):
\begin{itemize}
\item {\em $E/\Omega$ boundary:} As for the 2D Navier-Stokes turbulence considered before,
we have for this boundary separating the energy and the enstrophy cascades,
\begin{equation}
\label{1b}
k^2 \sim k_0^2,
\end{equation}
(i.e. a circle in the 2D $k$-space, $k_x^2 +k_y^2 = k_0^2.$)
\item {\em $E/Z$ boundary:} Equating the ratio of the energy density $|\omega_k|$
to the zonostropy density ${{k_x}^3}/{{k}^{8}}$
 to the initial value of this ratio, we get for the boundary separating the energy and the zonostrophy cascades,
\begin{equation}
\label{2b}
k^3/k_x \sim k_0^3/k_{0x}.
\end{equation}
\item {\em $\Omega/Z$ boundary:}  Equating the ratio of the enstrophy density $k_x$
to the zonostropy density ${{k_x}^3}/{{k}^{8}}$
 to the initial value of this ratio,  we get for the boundary separating the enstrophy and the zonostrophy cascades,
\begin{equation}
\label{3b}
k^4/k_x \sim k_0^4/k_{0x}.
\end{equation}
\end {itemize}

The first of these expressions, (\ref{1b}), says that (like in 2D turbulence before) the energy
must go to larger scales and the enstrophy must go to smaller scales.

The second expression, (\ref{2b}),  says that the energy must go to the zonal scales, $k_y \gg k_x$.
Moreover, this expression also poses a particular restriction on the path of the
energy to the zonal scales, e.g. for $k_y \gg k_x$ it should zonate at least as fast as
$k_y = \hbox{const} \; k_x^{1/3}$, see Fig.~(\ref{fig:triple_sketch}).

The last relation, (\ref{3b}), is also interesting. Because $k \ge k_x$, the curve (\ref{2b})
intersects the $k_x$-axis at a finite distance,
\begin{equation}
\label{kz-max}
k_x^* \sim k_0^{4/3} /k_{0x}^{1/3}.
\end{equation}
We see that the zonostrophy cannot cascade too far to large $k$'s, unless one starts with nearly zonal scales,
$k_{0y} \gg k_{0x}$. In particular, if $k_{0y} =k_{0x}$ we have $k_x^* = 2^{1/6} k_0$, i.e. the maximal
allowed wavenumber for the zonostrohy cascade is practically the same as the initial scale.
In other words, in this case the zonostrophy can only cascade to the larger scales.

\subsection{Alternative argument for zonation.}

The generalized Fj\o rtoft argument presented above was based on the zonostrophy
conservation which was proven for the wave kinetic equation and therefore, it is expected
to work well for the weakly nonlinear case. However, later we are going to present numerics
in which we "push" the formal  boundaries of validity and test the performance of the
triple cascade picture in the case when the initial nonlinearity is strong.

\begin{figure}[h!]
\begin{centering}
\includegraphics[width=0.65\textwidth]{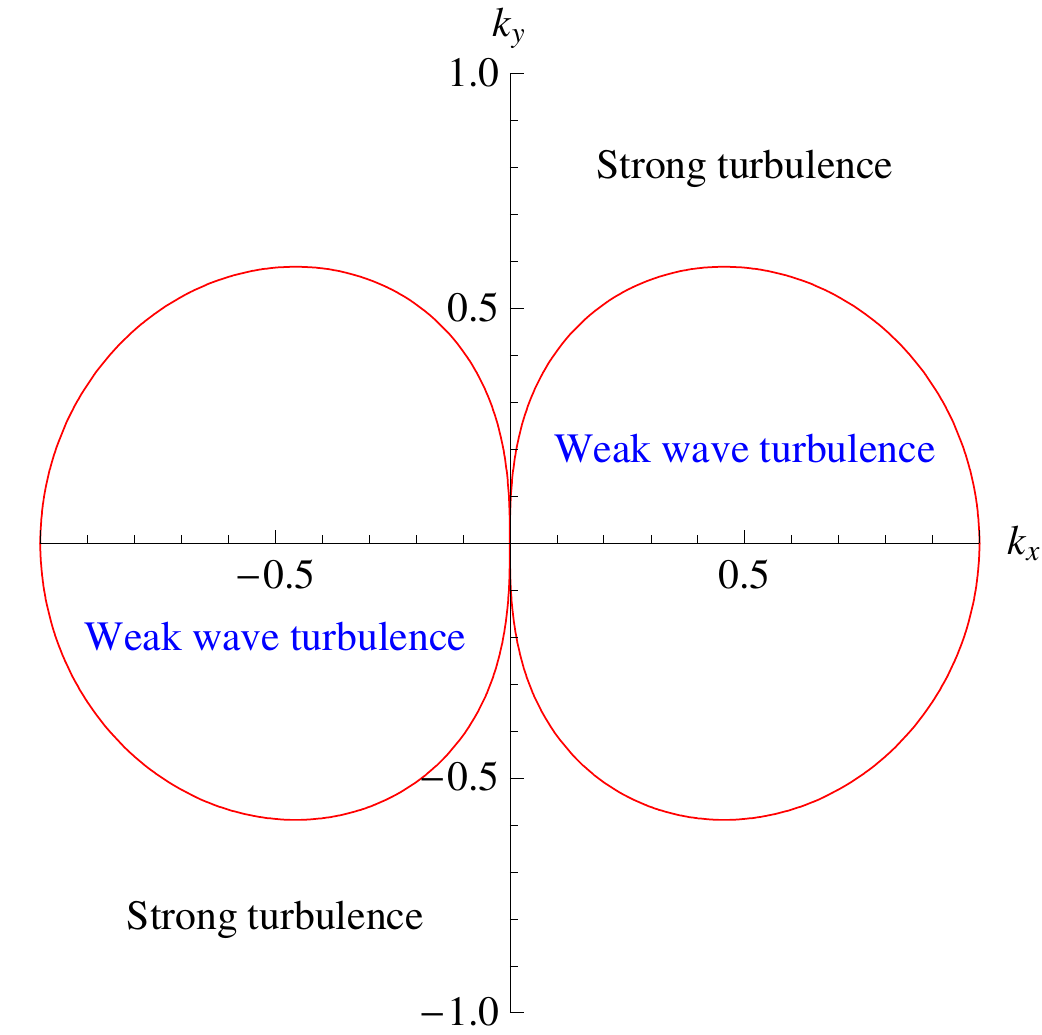}
\caption{\label{fig:lazy8} Dumb-bell curve in the Fourier space defining crossover from strong to weak turbulence.}
\end{centering}
\end{figure}

Because we are going to consider cases with strong nonlinearity, we would like
to mention an alternative argument for zonation which makes sense when the initial
turbulence is strong.
Suppose that turbulence is forced strongly at large $k$'s, so strongly that the linear term
 is negligible compared to the nonlinear one in the CHM equation (\ref{eq-CHM}).
 Since the linear term is the only source of anisotropy in the CHM model, the system is
 expected to build an isotropic inverse energy cascade identical to one of the 2D Navier-Stokes
 turbulence.
 As the inverse cascade progresses toward the larger scales, the linear
 dynamics ($\beta$-effect) become more and more important.
 A well-known boundary exists which defines the crossover from strong to weak turbulence,
 dominated by weakly nonlinear waves.
  This boundary is defined by the transitional wavenumber where the characteristic times
  of the linear and the nonlinear dynamics become equal.  As it follows from the
  above speculation about the inverse cascade, the scaling for the nonlinear time
  has to be taken from a Kolmogorov-type estimate, $\tau_{NL} \sim
  (\varepsilon k^2)^{-1/3}$ and the linear time is just the inverse wave frequency,
 $\tau_L = k^2/(\beta k_x)$. Equating  $\tau_{NL}$ and  $\tau_{L}$ we get
 \begin{equation}
\label{lazy8}
k^8 = k_{\beta}^5 k_x^3,
\end{equation}
where $k_{\beta} = ({\beta}^3/\varepsilon)^{1/5}$. This curve is
 plotted in Fig.~(\ref{fig:lazy8}) and it is
 the well-known
 `dumb-bell' or `lazy 8' curve  ~\cite{holloway,vallis}.
For the modes which lie outside the curve, isotropic turbulence is the dominant process, whereas for those modes inside the dumb-bell, anisotropic Rossby wave turbulence is the dominant process.

The alternative zonation argument is based on the observation that weak wave turbulence is much less
efficient in supporting the energy cascade than the strongly nonlinear interactions.
Thus one can suppose that the energy cascade does not penetrate
inside of the dumb-bell curve, but rather, it turns and continues along this curve to the larger scales.
This condition that the linear and the nonlinear times are balanced {\em scale-by-scale}
could be called a ``critical balance', in analogy with similar arguments about the
critical balance state discussed in MHD turbulence \cite{gs95}.

The critical balance  picture would mean that the energy cascade path of zonation would be
given by expression~(\ref{lazy8}), which incidentally is very close to
the path predicted by the Fj\o rtoft argument~(\ref{2b}), if $k_0 \sim k_\beta$
(the power $3/8$ indeed very close to  $1/3$).

A closer look reveals that the alternative zonation argument is actually not so ``alternative''
as it does not contradict our generalized Fj\o rtoft argument based on the zonostrophy
concentration. Indeed, if the energy followed a path through the scales at which the linear and the nonlinear
timescales balance then the zonostrophy would flow below this path, i.e. to the weakly nonlinear scales.
Thus, the zonostrophy would be supported at the weakly nonlinear scales and therefore would be conserved
(even though both the energy and the enstrophy would be supported in the intermediately and strongly
nonlinear scales in this case). But then all the three invariants are conserved (no need for weak nonlinearity
for conservation of $E$ and $\Omega$) and the generalized Fj\o rtoft argument is once again valid.

One should also keep in mind that  for the dumb-bell curve the estimation of the shape is
only approximate because it is based on the expression for $\tau_{NL}$ which
is valid, strictly speaking, only at $k$'s which are much greater than the dumb-bell (i.e.
in the region of the isotropic
inverse cascade). With this observation, we can put together the "kinematic"
view of  the Fj\o rtoft argument and the "dynamic" critical balance
condition, and suggest that:
\begin{itemize}
\item
Invariants $E$, $\Omega$ and $Z$ cascade in the sectors prescribed
by the generalized Fj\o rtoft argument, i.e. (\ref{1b}), (\ref{2b}) and (\ref{3b}).
\item
The energy cascade follows a path along the scales at which $\tau_{NL} \sim \tau_{L}$
which lies somewhere in the sector (\ref{2b}) and which does not strictly
follow the dumb-bell prescription  (\ref{lazy8}) (because of the unprecise definition
of the dumb-bell discussed above).
\item This picture is of course preceded by the usual isotropic inverse cascade
of $E$ in the case when the initial turbulence is very strong.
\end{itemize}

We emphasize that in this section we only considered the case when the initial
turbulence is strong. If the initial turbulence is weak, i.e. well within the
dumb-bell, then the energy cascade path remains in the weakly nonlinear scales
(at least for a while) rather than lie on the dumb-bell curve.

Now we will proceed to testing our theoretical predictions about the conservation
of the zonostrophy and the triple cascade behavior via  numerical
simulations of the CHM equation in the Fourier space, Eqn~(\ref{eq-CHMk}).

\section{NUMERICAL STUDY} \label{}

A pseudospectral code has been written to solve equation~(\ref{eq-CHMk}).  No dissipation is used and the initial condition is given by
\begin{equation}
\label{eq:ic}
\hat \psi_\textbf{k}|_{t=0}  = A e^{({\frac{|\textbf{k} - \textbf{k}_0|^2}{{k_*}^2} + i \phi_\textbf{k}})} +\hbox{image},
\end{equation}
where $k_0$ and $k_*$ are constants  and $\phi_\textbf{k}$ are random independent phases,
and by ``image'' we mean the mirror-reflected spectrum with respect to the $k_x$  axis.
Note that only the semi-plane $k_x \ge 0$ was used in our computations because
of the symmetry $\hat \psi_\textbf{-k} = \hat \psi_\textbf{k}^*$ arising from the fact that the streamfunction
$\psi$ in the CHM equation (\ref{eq-CHM}) is a real function.

We opted to simulate such an evolving non-dissipative system rather than a forced/dissipated
steady-state turbulence considered by the Fj\o rtoft argument because it appears to be more
physically relevant (since there appears to be no physically meaningful dissipation acting selectively
on nearly zonal and nearly meridional scales only).
Yet, we hope that the cascade picture obtained via the Fj\o rtoft argument is meaningful
for such decaying turbulence too, similar to what appears to be the case e.g. in the 2D Navier-Stokes
turbulence.

Of course, when calculating a non-dissipative system one has to be aware of the
possible bottleneck accumulation of turbulence near the maximum wavenumber after
the turbulent front reaches these scales. Thus we make sure to stop our simulations
before this happens.

\subsection{Centroids} \label{}
To quantify
 the cascades of the energy, enstrophy and zonostrophy in the
 time-evolving non-dissipative turbulence we introduce the {\em centroids}
  (``centres of mass'') of the total of each invariant defined respectively as follows,
\begin{equation}
\label{k-e}
\textbf{k}_E(t) = \frac{1}{E} \int  \textbf{k} \, k^2 |\hat \psi_k|^2 \, d\textbf{k},
\end{equation}

\begin{equation}
\label{k-om}
\textbf{k}_\Omega(t) = \frac{1}{\Omega} \int  \textbf{k} \, k^4 |\hat \psi_k|^2 \, d\textbf{k},
\end{equation}

\begin{equation}
\label{k-z}
\textbf{k}_Z(t) =  \frac{1}{Z} \int  \textbf{k} \, \frac{k_x^4}{k^6} (k_x^2 + 5 k_y^2) |\hat \psi_k|^2 \, d\textbf{k}.
\end{equation}

Of course, it is not {\em apriori} clear if the Fj\o rtoft argument formulated for the
steady-state forced/dissipated turbulence has a predictive power for the
trajectories of the centroids in the $k$-space.
In the present paper we ``experimentally'' verify that this is indeed the case.

For the 2D Navier-Stokes (Euler), it is actually possible to recast the Fj\o rtoft
argument for the non-dissipative evolving turbulence directly in terms of the
centroids in a rigorous way, see Appendix B.
However, the structure of the CHM is more involved and it is not clear if
one can produce a generalized
Fj\o rtoft's argument for the triple cascades in terms of the centroids in a
rigorous way. This is certainly an interesting question to be addressed in future.~\footnote{
We have been able to find some of the relevant inequalities in terms of the centroids,
but most of the conditions  restricting
the triple cascade sectors are still missing.}

  \subsection{Settings for the weakly nonlinear and the strongly nonlinear runs.}

  We have chosen two sets of parameters to be used in two runs corresponding to
  weak and strong initial nonlinearities respectively.

  The weakly nonlinear and the strongly nonlinear runs were performed at resolutions  $512^2$ and $1024^2$ respectively.
  This is because the weakly nonlinear systems evolve much slower that the strongly nonlinear ones
  and one has to compute them for much longer. Correspondingly,
  the centre of the initial spectrum and its width were chosen to be
  ${\bf k_0} = (20, 20)$ and $k_*=8$ in the weakly nonlinear run and
   ${\bf k_0} = (40, 40)$ and $k_*=16$
   in the strongly nonlinear run.

The initial amplitudes in the   weakly nonlinear and the strongly nonlinear runs
are $A=10^{-6}$ and  $A=5 \times 10^{-7}$ respectively.
Determining the relevant degree of nonlinearity $\sigma$ that corresponds to these initial
conditions is tricky.
One can directly estimate the linear and the nonlinear terms in the Eqn~(\ref{eq-CHMk}) and take into account the fact that there will be statistical
cancellations in the sum of the nonlinear term due to the random phases, i.e.
schematically
$$
\left|\sum_{j=1}^N \hbox{individual-term}_j \right| \sim \sqrt{N} \; \left|\hbox{individual-term} \right|.
$$
This way we get an estimate
\begin{equation}
\label{sigma}
\sigma \sim \frac{2 \sqrt{2 \pi} k_0^3 k_* A}{\beta},
\end{equation}
which gives $\sigma \sim 0.09$ for the weakly nonlinear run
and $\sigma \sim 0.7$ for the strongly nonlinear run.
However, it is likely that in the strongly nonlinear case the phases will quickly become
correlated to a certain degree. Evaluating the nonlinear term in the extreme
case when the phases are totally coherent would give an extra factor of $\sqrt{N} \sim
2 \sqrt{\pi k_*^2} \sim 100$, so we would have for the strongly nonlinear run
$\sigma \sim 70$, which  obviously is an overestimate.
From common sense, in a non-rigorous way, we believe that the true relevant value for $\sigma$
  in this case is closer to the random-phase estimate, but perhaps slightly higher,
  e.g. $\sigma \sim 1 - 2$, which is approximately ``on the dumb-bell''.

\subsection{Weakly Nonlinear Case} \label{}

\begin{figure}
\includegraphics[width=0.65\textwidth,angle=-90]{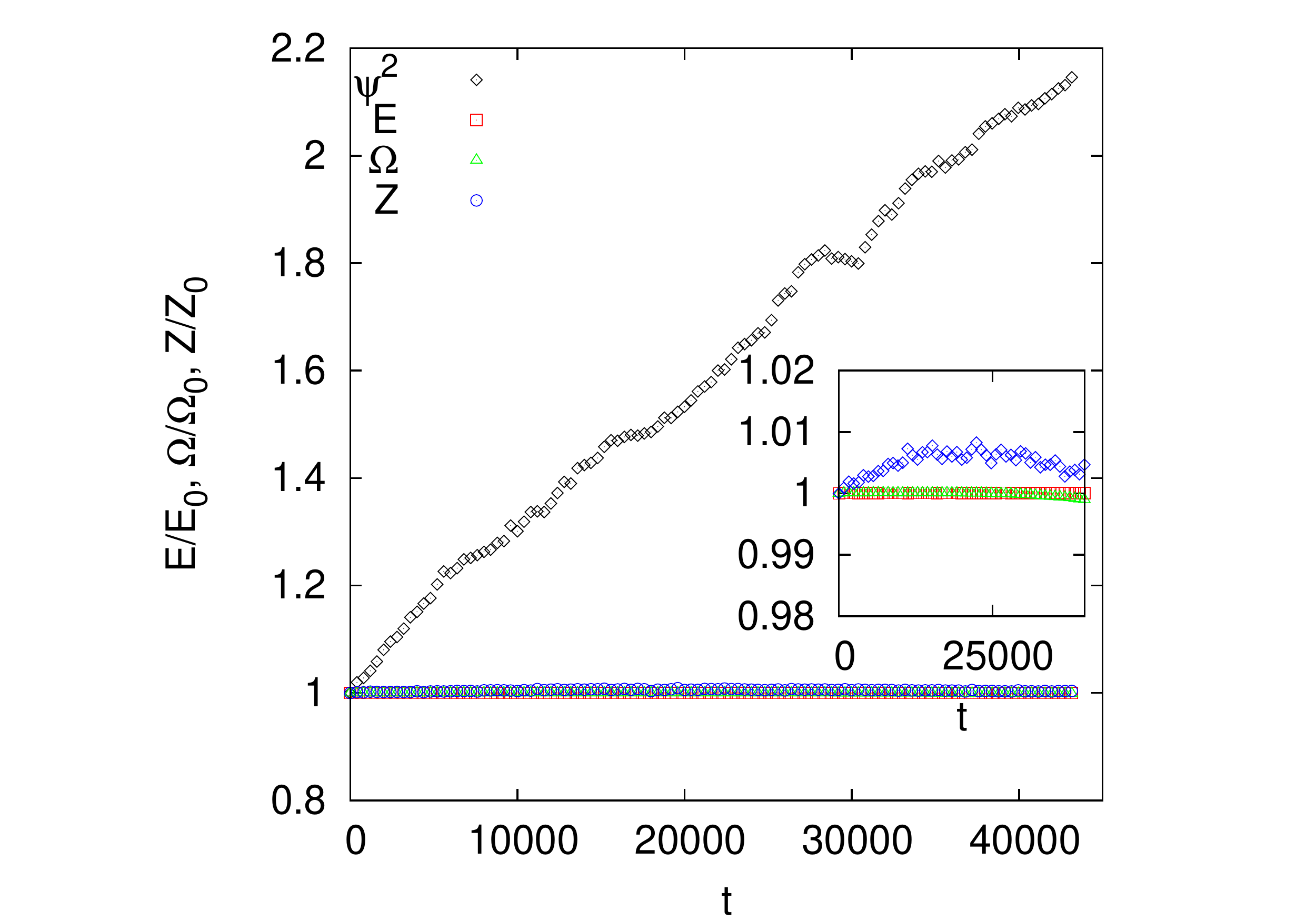}
\caption{\label{fig:cons_NL_low} Conservation of energy, enstrophy and zonostrophy for the weakly nonlinear case.  Non-conserved quantity $\Psi^2$ also shown.}
\end{figure}

For this case,  the degree of nonlinearity is weak, $\sigma \sim 0.09$,
 so that the initial turbulence is well within the dumb-bell.  Fig.~(\ref{fig:cons_NL_low}) shows the conservation of each invariant.  Because of the slow weakly nonlinear evolution, any quantity proportional to the turbulent intensity could look "conserved", so to demonstrate
 the true conservation of the zonostrophy we plot its time evolution along with a non-conserved quantity, $\int \psi^2 \, d{\bf x}$.
  It is clear that all three invariants, $E, \Omega$ and $Z$ are well conserved.
Namely, the energy is conserved within 0.01\%, the enstrophy - within 0.15\% and the zonostrophy is conserved within 1 \%.
Note that this is a first numerical demonstration of the conservation of the zonostrophy invariant.
We remind the reader that $Z$ is precisely conserved by the wave kinetic equation
(\ref{kineq})
and  therefore its conservation by the dynamical CHM equation~(\ref{eq-CHMk})
 is subject to the applicability conditions of this kinetic equation, namely the weak nonlinearity
and the random phases. It is not {\em apriori} clear how well these conditions
are satisfied throughout the $k$-space (particularly near the zonal scales).

\begin{figure}
\begin{centering}
\includegraphics[width=0.65\textwidth,angle=-90]{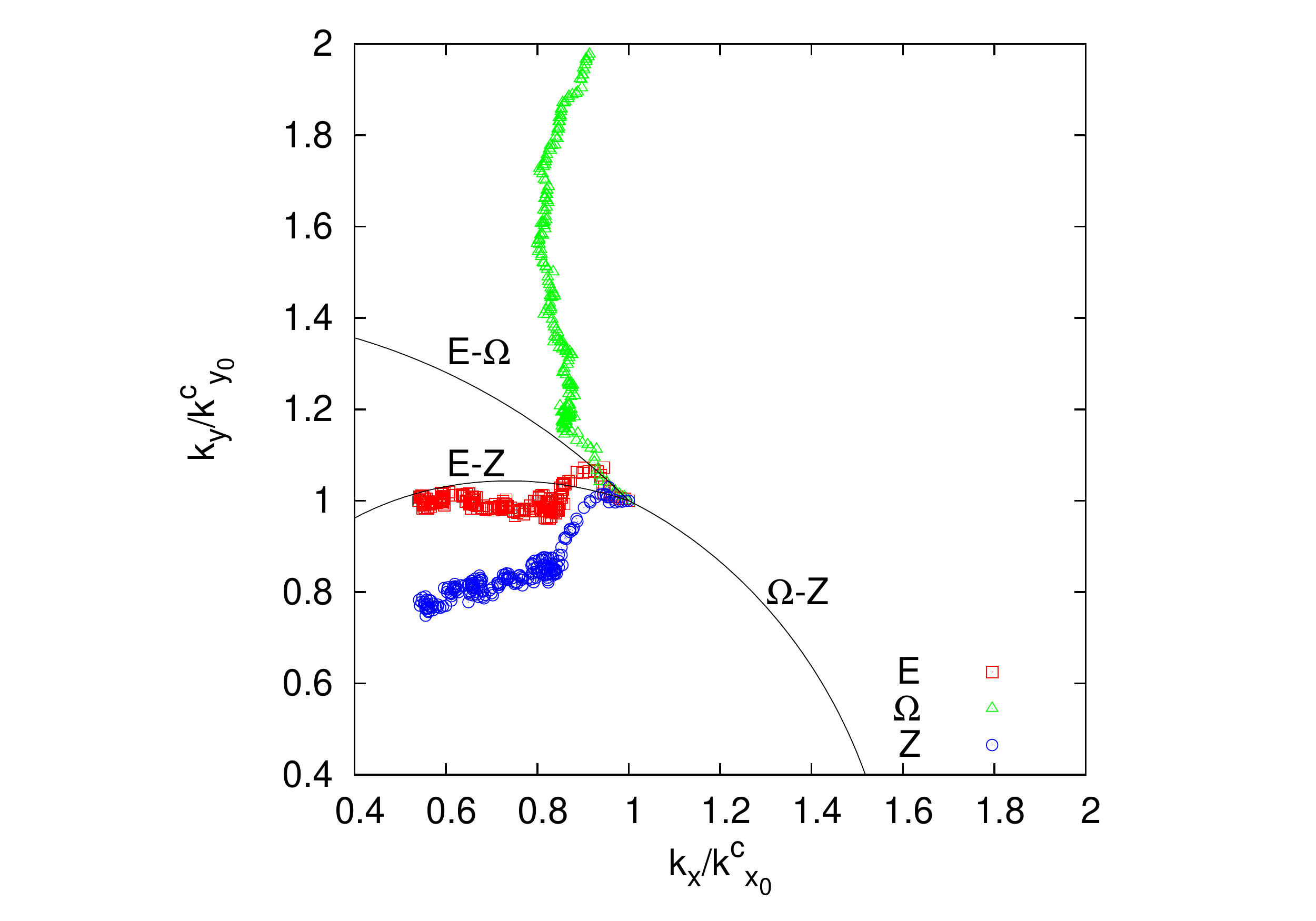}
\caption{\label{fig:cascade_NL_low} The cascades of energy, enstrophy and zonostrophy for the weakly nonlinear case, tracked by their centroids.}
\end{centering}
\end{figure}

The cascade directions for $E, \Omega$ and $Z$ are plotted in Fig.~(\ref{fig:cascade_NL_low})
 in terms of the paths followed by the respective centroids, (\ref{k-e}), (\ref{k-om}) and (\ref{k-z}).
 Note that for convenience  we normalize  the centroids to their initial values,
 $k_{E0}^c, k_{\Omega 0}^c$ and $k_{Z0}^c$ (which are different from $k_0$)
 so that the centroid paths start from the same point in Fig.~(\ref{fig:cascade_NL_low}).

 In Fig.~(\ref{fig:cascade_NL_low})
 we see that each invariant cascades well into its predicted sector.
 Interestingly, the enstrophy and the zonostrophy paths are well inside their
 respective cascade sectors, whereas the energy follows the boundary of its sector with the zonostrophy sector. One should remember, however, that the boundaries between the sectors are
 not sharp because the Fj\o rtoft argument operates with strong inequalities ($\ll$ and $\gg$ rather
 than $<$ and $>$).

\begin{figure*}
\begin{centering}
\includegraphics[width=1.0\textwidth]{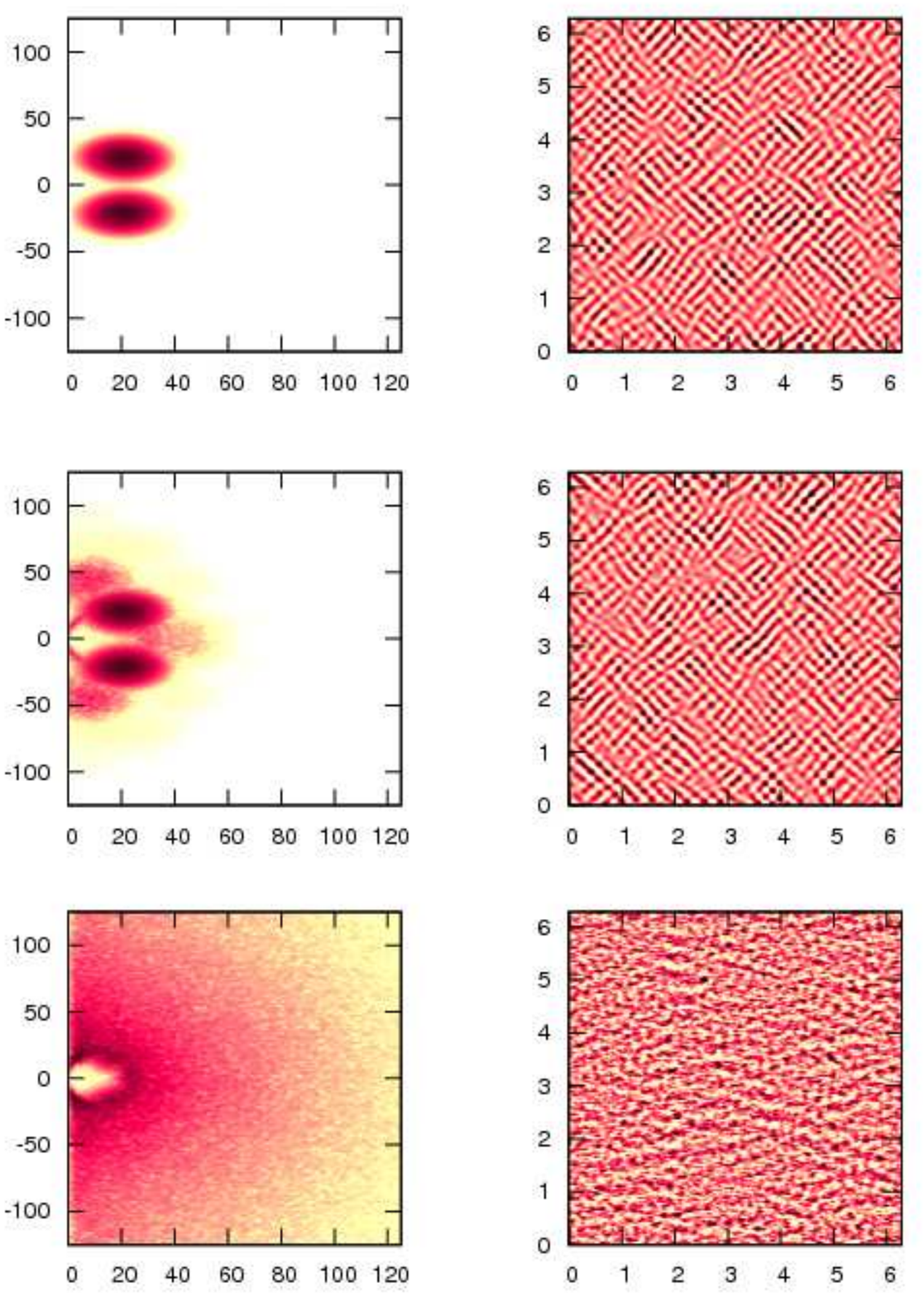}
\caption{\label{fig:contours_NL_weak} Contour plots of the $2D$ energy spectrum in $k$-space (left) and corresponding contours of vorticity in $x$-space (right) for the weakly nonlinear case.}
\end{centering}
\end{figure*}

Three successive frames of the energy spectrum in the 2D $k$-space along with the $x$-space frames of vorticity distributions at the same moments of time
for the weak nonlinearity case are shown in Fig.~(\ref{fig:contours_NL_weak}).
The initial spectrum, which represents a  gaussian spot centered at ${\bf k_0}$ and its mirror image, is seen
to grow "arms" toward the coordinate origin, so that a closed "ring" forms and then starts shrinking in size.
The ring is suggestive of the dumb-bell shape (when complemented with the other half of the distribution
at $k_x<0$), although the similarity is only visual  rather than quantitative, because
the nonlinearity is quite small.
Presumably, the growing of arms and the ring shrinking  are indicative of the structure of the anisotropic inverse energy
cascade process.
On the respective vorticity $x$-plots, we see initial dominant shortwave components propagating
at $\pm 45^o$ (corresponding to the position of the initial maxima in the spectrum) which in time
evolve into a more disordered turbulent state with a predominant zonal orientation.

\subsection{Strongly Nonlinear case} \label{}

\begin{figure}
\begin{centering}
\includegraphics[width=0.65\textwidth,angle=-90]{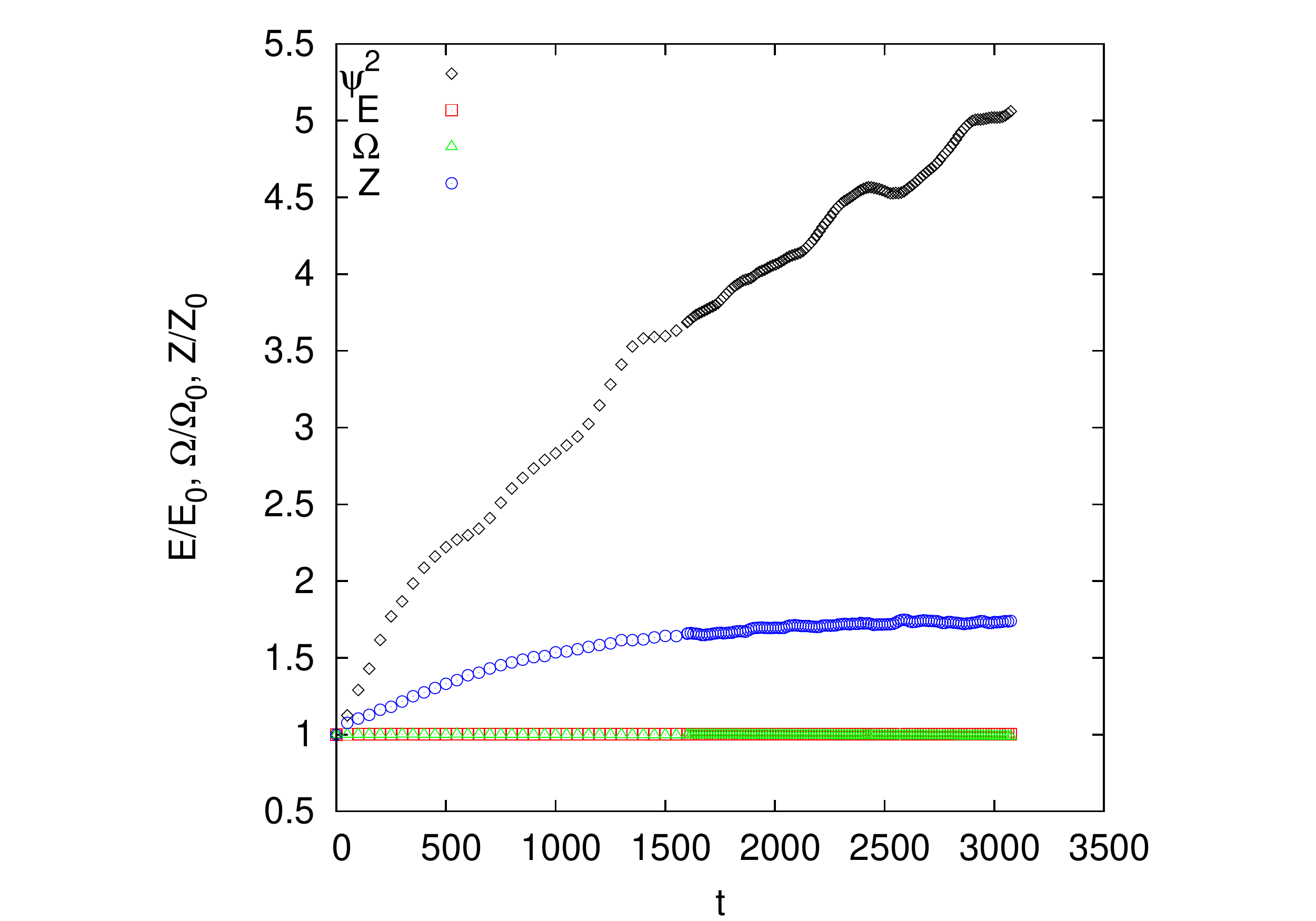}
\caption{\label{fig:cons_NL_high} Conservation of energy, enstrophy and zonostrophy for the strongly nonlinear case.  Non-conserved quantity $\Psi^2$ also shown.}
\end{centering}
\end{figure}

For this case, $\sigma \sim 1 - 2$ so that the initial turbulence is near the boundary of the dumb-bell.  Fig.~(\ref{fig:cons_NL_high}) shows the conservation of each invariant.  While the energy and enstrophy are still well conserved (the energy within 0.2\% and the
enstrophy - within 1.2\%), the zonostrophy is not conserved initially.
This is not surprising considering that the zonostrophy is only expected to be conserved if the nonlinearity is weak.
What is more interesting, however, is that the zonostrophy growth saturates as time proceeds, so that
the zonostrophy is rather well conserved in this case for large times. This suggests, as we argued before, that
for large times the scales that support the zonostrophy invariant are weakly nonlinear, even though
the energy scales probably remain moderately nonlinear, and the enstrophy scales are definitely
strongly nonlinear.

\begin{figure}
\begin{centering}
\includegraphics[width=0.65\textwidth,angle=-90]{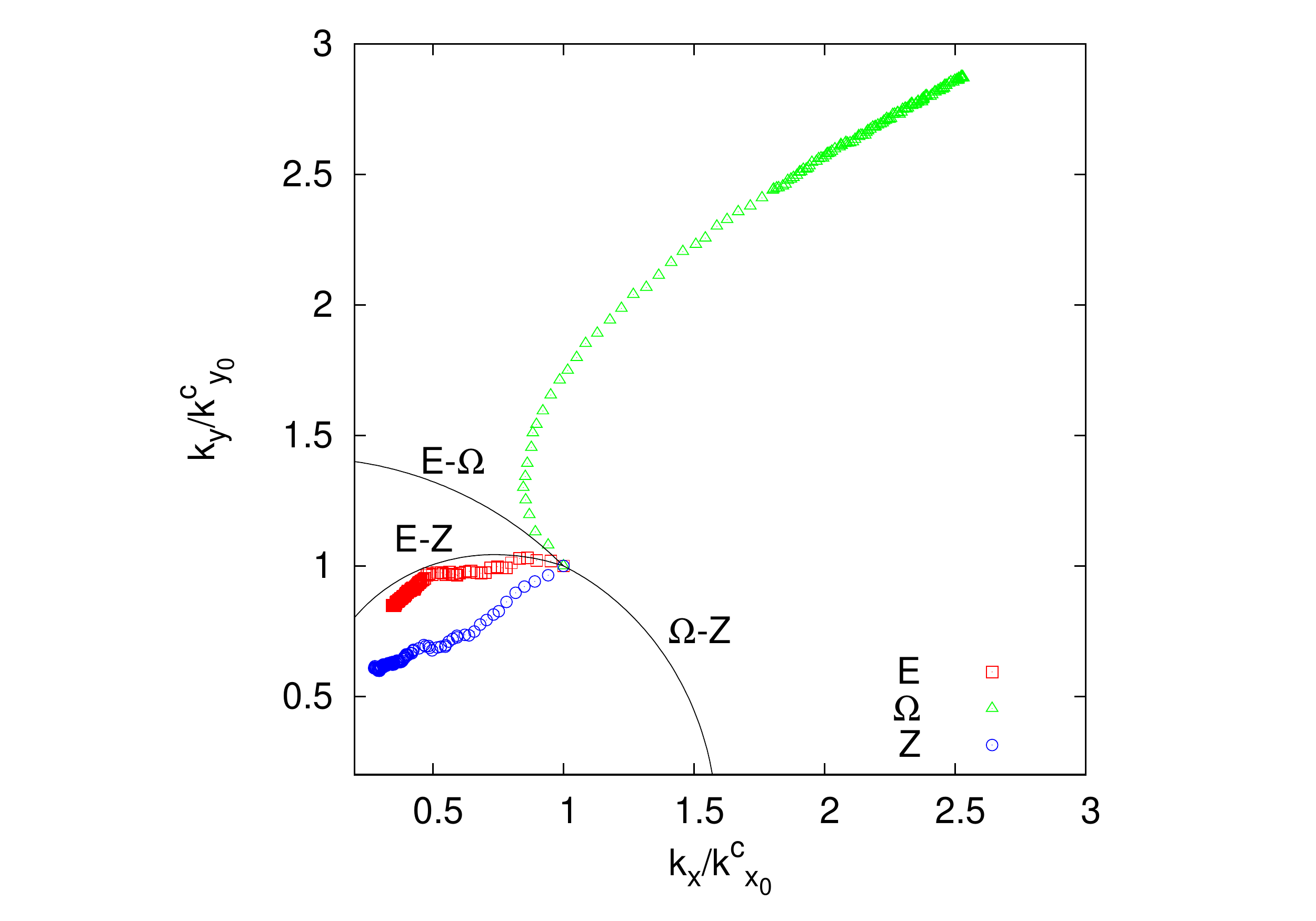}
\caption{\label{fig:cascade_NL_high} The cascades of energy, enstrophy and zonostrophy for the strongly nonlinear case, tracked by their centroids.}
\end{centering}
\end{figure}

The cascade paths for $E, \Omega$ and $Z$ in terms of the respective centroids are plotted in Fig.~(\ref{fig:cascade_NL_high}).
  Once again we see a picture which is similar to the one already observed for the weakly nonlinear case:
  the enstrophy and the zonostrophy cascades lie well inside their respective theoretically predicted
  sectors, and the energy cascade  follows the boundary of its sector. It is quite possible that the energy path lies
  in the ``critically balanced'' scales where the nonlinear and the linear time scales are of the same order.
  However the measurement of the nonlinear time scale is quite ambiguous and it is still unclear if the
  critical balance approach can be formulated in a more precise way in this case.
  In any case, one can clearly see that even in this strongly nonlinear case the zonostrophy invariant
  is conserved for large times, and that the triple cascade picture predicted using this invariant
  provides a reasonable description of the turbulence evolution and explanation of the zonal jet formation.

\begin{figure*}
\begin{centering}
\includegraphics[width=1.0\textwidth]{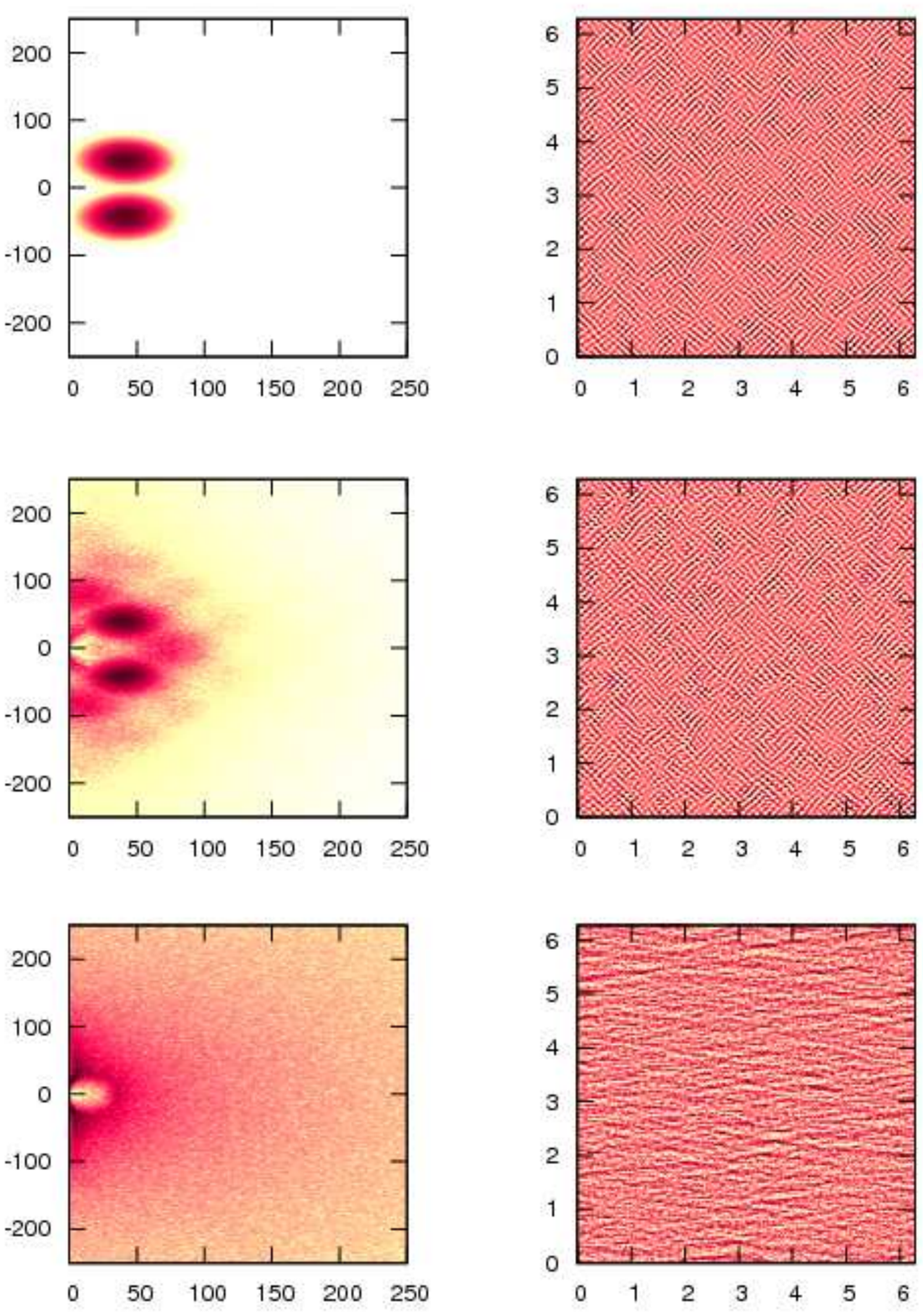}
\caption{\label{fig:contours_NL_strong}Contour plots of the $2D$ energy spectrum in $k$-space (left) and corresponding contours of vorticity in $x$-space (right) for the strongly nonlinear case.}
\end{centering}
\end{figure*}

We have also looked at the energy spectra in the 2D wavenumber space and at the
2D vorticity distributions on the $x$-plane evolving in time, see Fig.~(\ref{fig:contours_NL_strong}).
The essential features of the evolution of these distributions appeared to
be remarkably similar to the ones of the weakly nonlinear case, with somewhat more evident
zonal jets at later times. This can be explained by the fact that the strongly nonlinear systems
evolve faster than the weakly nonlinear ones, so that what we see here is a more advanced
stage of zonation.

\section{SUMMARY} \label{}
In the present paper the generalized Fj\o rtoft argument was used to predict a triple cascade behaviour
of the CHM turbulence, in which the energy, the enstrophy and the zonostrophy are
cascading into their respective non-intersecting sectors in the scale space.
These cascades are anisotropic and the energy cascade is predicted to be
directed to the zonal scales, which provides a physical explanation and the character
of the formation of the zonal  jets in such systems.

The zonostrophy conservation, as well as the triple cascade picture, were tested numerically
for the cases of both weak and strong initial nonlinearities.
The zonostrophy invariant was shown to be well conserved in the weakly nonlinear case.
Moreover, the zonostrophy conservation was also observed for the case with strong
initial nonlinearity after a transient non-conservative time interval.
Presumably, this is because the zonostrophy moves in time to the scales that are
weakly nonlinear (even though the energy and the enstrophy remain in the strongly
nonlinear parts of the Fourier space).
Using the energy, the enstrophy and the zonostrophy centroids for tracking the
transfers of these invariants in the Fourier space, we demonstrated that
all the three invariants cascade as prescribed by the triple cascade Fj\o rtoft argument
in both the weakly nonlinear and in the strongly nonlinear cases.
The energy appears to be somewhat special among the three invariants in that
it tends to cascade along the edge of the sector allowed by the Fj\o rtoft argument,
namely along the curve
$k \propto k_x^{1/3}$.

We believe that further studies would be helpful, both theoretical and numerical, for
establishing the conditions under which the zonostrophy is conserved,
in particular finding out the extent to which the statistical properties
of the system (e.g. random phases) are
important in addition to weak nonlinearity of the zonostrophy supporting scales.
It would be also interesting to study the behaviour of zonostrophy in the other
setups within the CHM model, e.g. the modulational instability
and truncated systems of coupled resonant triads.


\begin{acknowledgement}

We acknowledge stimulating discussions with Alexander Balk, Peter Bartello, David Dritschel,
 Volker Naulin and Peter Rhines, whose comments and suggestions we greatly
appreciate. Our special thanks to Colm Connaughton, who helped developing the numerical code,
and to Gregory Eyink, who proposed using Cauchy-Schwartz inequality for proving
the inequalities involving the centroids.

\end{acknowledgement}

\section*{Appendix A - Small-scale limit of  Zonostrophy}

The general expression for the zonostrophy density is given by ~\cite{balk-gen}:

\begin{equation}
\label{eq:arctan}
\zeta = \arctan {\frac{k_y-k_x \sqrt{3}}{\rho k^2}} - \arctan {\frac{k_y+k_x \sqrt{3}}{\rho k^2}}
\end{equation}
Expanding this expression in the powers of $1/\rho$ up to ninth order, we get
\begin{eqnarray*}
\label{eq:taylor_Z}
\zeta = && -2 \sqrt{3} \frac{k_x }{\rho k^2} + 2 \sqrt{3} \frac{k_x}{\rho^3 k^4} \\
&& - 2 \sqrt{3} \frac{k_x }{\rho^5 k^{10}} [k_y^4+6 k_x^2 k_y^2 +\frac{9}{5} k_x^4] \\
&& + 2 \sqrt{3} \frac{k_x }{\rho^7 k^{14}} [\frac{27}{7} k_x^6 + 27 k_x^4 k_y^2 + 15k_x^2 k_y^4 + k_y^6] \\
&& - 2 \sqrt{3} \frac{k_x }{\rho^9 k^{18}} [ 9 k_x^8 + 108 k_x^6 k_y^2 + 126k_x^4 k_y^4 \\
&& + 28k_x^2 k_y^6 + k_y^8] + O(\rho^{-10}).
\end{eqnarray*}
We note that the Taylor expansion of the frequency $\omega_k = -\beta/(k^2 + \rho^{-2})$ in
 the powers of $1/\rho$  is
\begin{equation}
\label{eq:omega_n}
\omega_k = \sum_{n=1}^\infty \omega^{(n)} = \sum_{n=1}^\infty \frac{(-1) ^n \beta \rho^2 k_x}{ (\rho k)^{2n}}.
\end{equation}
We see that in the leading order $\zeta  = \frac{2 \sqrt{3}}{\beta \rho} \omega_k$, i.e.  in the small-scale
limit $\zeta$ is proportional to the energy and not an independent invariant.

Thus, to find the truly independent invariant in the small-scale limit
we must subtract this "energy" part. To have a simpler expression
which is $\rho$-independent in the leading order, we will also multiply the result by
$-\frac{5 \rho^5}{8 \sqrt{3}}$.
Re-defined this way, the zonostrophy invariant is
\begin{eqnarray*}
\label{eq:omegas}
\tilde \zeta =
-\frac{5 \rho^5}{8 \sqrt{3}} (\zeta - 2 \sqrt{3}  \omega/\beta \rho) =\hspace{6cm}\\
	k_x^3 (\frac{5 k_y^2+k_x^2} { k^{10}} - 5\frac{\frac{5}{7}k_x^4 + 6k_x^2 k_y^2 + 3k_y^4} {\rho^2 k^{14}} +
	 5\frac{2k_x^8 + 26k_x^6 k_y^2 + 30k_x^4k_y^4 +6k_x^2 k_y^2} {\rho^4 k^{18}}),
\end{eqnarray*}
which in the limit $\rho \to \infty$ becomes the expression we were looking for,
\begin{equation}
\label{eq:zonostrophy}
\tilde \zeta = k_x^3 \frac{k_x^2+5 k_y^2} {k^{10}}.
\end{equation}

\section*{Appendix B - Fj\o rtoft argument in terms of the centroids.}

Let us consider an evolving hydrodynamic 2D turbulence in the absence of
forcing and dissipation.
Here we will re-formulate the Fj{\o}rtoft argument
in terms of the energy and enstrophy centroids in the ${k}$ and ${l}$ (i.e. scale) spaces.
This formulation will be rigorous and quite useful for visualizing the directions of transfer of
of the energy and the enstrophy. In contrast with the  version of the Fj{\o}rtoft argument give in
the main text, this formulation is for
a non-dissipative turbulence rather than a forced/dissipated system.

The energy and the enstrophy $k$-centroids are defined respectively as
\begin{eqnarray}
\label{ener-k-centroid}
k_E  &=& \int_0^\infty k \, E_k \, dk /E, \\
\label{enst-k-centroid}
k_\Omega &=&  \int_0^\infty k^3 \, E_k \, dk /\Omega,
\end{eqnarray}
and the energy and the enstrophy $l$-centroids as defined respectively as
\begin{eqnarray}
\label{ener-l-centroid}
l_E &=& \int_0^\infty k^{-1} \, E_k \, dk /E, \\
\label{enst-l-centroid}
l_\Omega &=&  \int_0^\infty k \, E_k \, dk /\Omega \equiv k_E E/\Omega,
\end{eqnarray}
where $E_k$ is the 1D energy spectrum (i.e. the energy density in $|k|$).

\begin{theorem}
Assuming that the integrals defining $E,\Omega, k_E,k_\Omega, l_E$ and $l_\Omega$ converge, the following
inequalities hold,
\begin{eqnarray}
\label{k_e-ineq}
k_E &\le& \sqrt{\Omega/E}, \\
\label{k_om-ineq}
k_\Omega &\ge& \sqrt{\Omega/E}, \\
\label{k_om-k_e}
k_E k_\Omega &\ge& {\Omega/E}, \\
\label{l_e-ineq}
l_E &\ge& \sqrt{E/\Omega}, \\
\label{l_om-ineq}
l_\Omega &\le& \sqrt{E/\Omega}, \\
\label{l_e-l_om}
l_E l_\Omega &\ge& {E/\Omega}.
\end{eqnarray}
\end{theorem}

We are going to prove this theorem using Cauchy-Schwartz inequality~\footnote{The suggestion to use
the Cauchy-Schwartz inequality for reformulating the Fj\o rtoft argument in terms of the centroids was
made to us by Gregory Eyink during the INI workshop the proceedings of which are published in this book.},
which states that
$$
\left|\int_0^\infty f(k) g(k) \, dk \right| \le \left| \int_0^\infty f^2(k)  \, dk \right|^{1/2} \left| \int_0^\infty  g^2(k) \, dk \right|^{1/2}
$$
for any functions $f(k), g(k) \in {L}^2 $. We will only deal with positive functions, so the
absolute value brackets may be omitted. Being in $L^2$ in our case means that all the relevant
integrals converge, as suggested in the statement of the problem.

First, let us consider integral $\int k E\, dk$ and apply Cauchy-Schwartz inequality as follows,
$$
\int_0^\infty k E\, dk = \int_0^\infty (k E^{1/2})(E^{1/2}) \, dk \le \left( \int_0^\infty k^2 E  \, dk \right)^{1/2} \left( \int_0^\infty  E \, dk \right)^{1/2},
$$
which immediately yields (\ref{k_e-ineq}) and (\ref{l_om-ineq}).

Second, let us consider $\Omega = \int k^2 E\, dk$ and split it as,
$$
\int_0^\infty k^2 E\, dk = \int_0^\infty (k^{3/2} E^{1/2})(k^{1/2} E^{1/2}) \, dk
\le \left( \int_0^\infty k^3 E  \, dk \right)^{1/2} \left( \int_0^\infty  k E \, dk \right)^{1/2},
$$
which immediately yields (\ref{k_om-k_e}). Combining (\ref{k_om-k_e}) with (\ref{k_e-ineq})
gives (\ref{k_om-ineq}).

Now, let us split  $\int k E\, dk$ in a different way,
$$
\int_0^\infty k E\, dk = \int_0^\infty (k^{3/2} E^{1/2})(k^{-1/2} E^{1/2}) \, dk
\le \left( \int_0^\infty k^3 E  \, dk \right)^{1/2} \left( \int_0^\infty  k^{-1} E \, dk \right)^{1/2},
$$
which immediately yields (\ref{l_e-l_om}). Combining (\ref{l_e-l_om}) with (\ref{l_om-ineq})
gives (\ref{l_e-ineq}).

We see that according to inequalities (\ref{k_e-ineq}) and (\ref{k_om-ineq}), during
the system's evolution the energy
centroid $k_E(t)$ is bounded from above and the enstrophy centroid
$k_\Omega(t)$ is bounded from below (both by the same wavenumber $k=\sqrt{\Omega/E}$),
as one would expect from Fj{\o}rtoft argument.
Further, inequality (\ref{k_om-k_e}) means that if $k_E(t)$ happened to move to small
$k$'s then $k_\Omega(t)$ \textit{must} move to large $k$'s, that is roughly, there cannot be
inverse cascade of energy without a forward cascade of enstrophy.
Note that there is no complimentary restriction which would oblige $k_E(t)$ to become
small when $k_\Omega(t)$ goes large, so the $k$-centriod part of the Fj{\o}rtoft argument
is asymmetric, and one has to consider the $l$-centroids to make it symmetric.
Indeed, in additions to conditions (\ref{l_e-ineq}) and (\ref{l_om-ineq}) which are similar to
(\ref{k_e-ineq}) and (\ref{k_om-ineq}), we have inequality (\ref{l_e-l_om}) meaning
 that if $l_\Omega(t)$ happened to move to small
$l$'s then $l_E(t)$ \textit{must} move to large $l$'s, i.e. any forward cascade of enstrophy
must be accompanied by an inverse cascade of energy.

Importantly, we do not always have $k_E \sim 1/l_E$ and $k_\Omega \sim 1/l_\Omega$.
Indeed, consider a state with spectrum $E_k \sim k^{-5/3}$ for $k_a < k < k_b$ (with $k_b \gg k_a$)
and $E_k \equiv 0$
outside of this range. It is easy to see  that for this state $k_E \sim k_b^{1/3} k_a^{2/3}$ and $l_E \sim 1/k_b$ (i.e. $k_E \nsim 1/l_E$)
and $k_\Omega \sim 1/l_\Omega \sim k_b$.

\end{document}